\documentclass[%
 reprint,
 amsmath,amssymb,
 aps,
]{revtex4-1}

\usepackage{graphicx}
\usepackage{dcolumn}
\usepackage{bm}
\usepackage{threeparttable}
\usepackage{multirow}

\begin{document}

\preprint{APS/OO-AP1roG}

\title{Projected seniority-two orbital optimization of the antisymmetric product of one-reference orbital geminal}

\author{Katharina Boguslawski}
\author{Pawe{\l} Tecmer}%
\author{Peter A. Limacher}
\author{Paul A. Johnson} 
\author{Paul W. Ayers} 
\email{ayers@mcmaster.ca}
\affiliation{%
 Department of Chemistry and Chemical Biology, McMaster University, Hamilton, 1280 Main Street West, L8S 4M1, Canada 
}%
\author{Patrick Bultinck}
\affiliation{%
Department of Inorganic and Physical Chemistry, Ghent University, Krijgslaan 281 (S3), 9000 Gent, Belgium
 \\}
 \vspace{-2cm}
\author{Stijn De Baerdemacker}
\author{Dimitri Van Neck}
\affiliation{%
Center for Molecular Modelling, Ghent University, Technologiepark 903, 9052 Gent, Belgium
 \\}

\date{\today}

\begin{abstract}
We present a new, non-variational orbital-optimization scheme for the Antisymmetric Product of one-reference orbital Geminal wave function. Our approach is motivated by the observation that an orbital-optimized seniority-zero configuration interaction (CI) expansion yields similar results to an orbital-optimized seniority-zero-plus-two CI expansion [J. Chem. Phys., 135, 044119 (2011)]. A numerical analysis is performed for the C$_2$, LiF and CH$_2$ molecules as well as for the symmetric stretching of hypothetical (linear) hydrogen chains. For these test cases, the proposed orbital-optimization protocol yields similar results to its variational orbital optimization counterpart, but prevents symmetry-breaking of molecular orbitals in most cases.
\end{abstract}

\maketitle


\section{Introduction}
Two-electron functions, also called geminals, were introduced in quantum chemistry many decades ago \cite{Hurley_1953,Kutzelnigg_1964,Coleman_1965,Kutzelnigg_1965,Miller1968,Ortiz_1981,Surjan_1999} and can be considered as a generalization of one-particle functions (orbitals) \cite{Surjan_2012,Kutzelnigg2012}. In the geminal model, the fundamental building blocks for the electronic wave function are electron pair states: in contrast to Slater determinants, an antisymmetric product of (occupied) orbitals, the wave function is constructed as an antisymmetric product of geminals \cite{Hurley_1953,McWeeny1963}. A desirable feature of geminal-based methods is that electron correlation effects are built into the electronic wave function by construction. As a result, geminal-based approaches are well-suited to describe strong (static and nondynamic) electron correlation effects \cite{Cassam-Chenai2006,Cassam-Chenai2007,scuseria2009constrained,scuseria2011projected,Surjan_2012,Kutzelnigg2012,Limacher_2013,Johnson_2013,ellis2013pair,pawel_jpca-2014,p-CCD,Tamar-p-CCD}.

In the general case, all orbitals are allowed to contribute to each geminal resulting in the antisymmetric product of interacting geminals~\cite{Bratoz1965,Coleman_1965,Silver_1969,Silver_1970,APIG-1,APIG-2,Surjan-bond-1984,Surjan-bond-1985,Surjan-bond-1994,Surjan-bond-1995,Surjan-bond-2000,Surjan_1999,Rosta2002,Surjan_2012} (APIGs). In particular, the APIG model represents an excellent parameterization of the doubly occupied configuration interaction~\cite{DOCI} (DOCI) wave function. Although this ansatz permits us to correlate all orbital pairs, it is computationally intractable. Thus, different restrictions on the orbital pairing scheme have been introduced that are computationally feasible in practical applications. Examples are the antisymmetric product of strongly orthogonal geminals (APSG) and generalized-valence-bond perfect-pairing \cite{Hurley_1953,Parks_1958,Goddard1972,Goddard1973} (GVB-PP) methods. However, the severe restrictions in the orbital pairing scheme neglects correlation effects between orbitals assigned to different geminals.

Recently, we have presented new geminal-based wave function ans\"atze that approximate the APIG wave function, but are computationally tractable~\cite{Johnson_2013,Limacher_2013}. One promising approach represents the antisymmetric product of 1-reference orbital geminal (AP1roG)~\cite{Limacher_2013}. Specifically, the AP1roG model represents an efficient parameterization of the DOCI wave function, but requires only mean-field computational cost. In contrast, traditional DOCI implementations suffer from factorial scaling. The AP1roG wave function ansatz can be written in terms of one-particle functions as a fully general pair-coupled-cluster-doubles wave function \cite{p-CCD}, i.e.,
\begin{equation}\label{eq:ap1rog}
|{\rm AP1roG}\rangle = \exp \left (  \sum_{i=1}^P \sum_{a=P+1}^K c_i^a a_a^{\dagger}  a_{\bar{a}}^{\dagger}a_{\bar{i}} a_{i}  \right )|\Phi_0 \rangle,
\end{equation}
where $a_{p}^{\dagger}$, $a_{\bar{p}}^{\dagger}$ and $a_{p}$, $a_{\bar{p}}$ are the electron creation and annihilation operators for $\alpha$- ($p$) and $\beta$-electrons ($\bar{p}$), and $|\Phi_0 \rangle$ is some independent-particle wave function (usually the Hartree--Fock (HF) determinant). 
In the above equation, we used the standard notation where indices $i$ and $a$ correspond to occupied and virtual orbitals with respect to $|\Phi_0 \rangle$. $P$ denotes the number of electron pairs ($P=N/2$ with $N$ being the total number of electrons) and $K$ is the number of one-particle functions. $\{c_i^a\}$ are the geminal coefficients and link the geminal wave function with the underlying one-particle basis functions \cite{Surjan_2012}. Especially, the geminal coefficient matrix encodes the orbital-pairing scheme in the geminal wave function. We should emphasize that, in the AP1roG ansatz, all virtual orbitals are allowed to contribute to each geminal. Thus, unlike APSG and GVB-PP, our approach does not require the orbital-pairing scheme to be optimized~\cite{Rassolov-2002}. 

In order to ensure size-consistency, we have to optimize the one-particle basis functions~\cite{Limacher_2013}. This can be done in a fully variational manner~\cite{OO-AP1roG}, analogous to orbital-optimized coupled cluster~\cite{Helgaker_book,Scuseria1987,Kohn2005,Ugur_2011} (OCC). The first variational orbital optimization of the AP1roG wave function was recently presented by some of us~\cite{OO-AP1roG,Piotrus_Mol-Phys}. The orbitals are then chosen to minimize the AP1roG energy expression subject to the constraint that the geminal coefficient equations (\emph{cf.} ref.~\cite{Limacher_2013}) are satisfied. The energy Lagrangian takes thus the form
\begin{align}\label{eq:voo}
\mathcal{L} = &\langle \Phi_0 |  e^{- \bm \kappa} \hat{H} e^{\bm \kappa} | {\rm AP1roG} \rangle + \nonumber \\
              &\sum_{i,a} \lambda_i^a \big( \langle \Phi_{i \bar{i}}^{a \bar{a}} | e^{- \bm \kappa} \hat{H} e^{\bm \kappa} |
              {\rm AP1roG} \rangle - Ec_i^a \big),
\end{align}
with $\{\lambda_i^a\}$ being the Lagrange multipliers. In the above equation, $\bm \kappa$ is the generator of orbital rotations
\begin{equation}\label{eq:kappa}
\bm{\kappa} = \sum_{p>q} \kappa_{pq} (a^\dagger_p a_q - a^\dagger_q a_p),
\end{equation}
where $(\kappa_{pq})$ is a skew-symmetric matrix, and transforms into a new orthogonal basis with a transformation $U=e^{\bm \kappa}$ and $ e^{- \bm \kappa} \hat{H} e^{\bm \kappa} $ being the Hamiltonian in the rotated basis~\cite{Helgaker_book}. $\Phi_{i \bar{i}}^{a \bar{a}}$ denotes a single-pair-excited determinant with respect to the reference determinant $\Phi_0$. The variational orbital gradient is obtained as the derivative of the Lagrange energy functional with respect to the orbital rotation coefficients $\{\kappa_{pq}\}$, $\partial \mathcal{L}/\partial \kappa_{pq}$. We should note that, in contrast to OCC theory, occupied--occupied and virtual--virtual orbital rotations are non-redundant and have to be considered in the orbital gradient equations~\cite{OO-AP1roG}. Thus, the indices $p$ and $q$ run over all occupied and virtual orbitals. The Lagrange multipliers $\{\lambda_i^a\}$ require the solution of an additional set of equations defined by $\partial \mathcal{L}/\partial \lambda_i^a$. 
In the following, the variationally orbital optimized AP1roG approach is labeled as vOO-AP1roG. 

\section{Decoupling of the Seniority-Two Sector}
In this article, we present an alternative approach to optimize the one-particle basis functions that is not guided by the variational principle. Our approach is inspired by the work of Bytautas \emph{et al.}~\cite{Bytautas2011} where CI-expansions restricted to the seniority-zero ($\Omega = 0$) and seniority-zero-plus-two sectors ($\Omega=0,2$) yield similar results if symmetry-adapted optimized molecular orbitals are used. The seniority number $\Omega$ is defined as the number of unpaired electrons in a Slater determinant~\cite{Bytautas2011,Alcoba2013}, \emph{i.e.}, the number of singly-occupied orbitals. It can be used as an alternative partitioning of the Hilbert space: instead of grouping Slater determinants according to their excitation (or substitution) level with respect to some reference determinant, Slater determinants can be grouped according to their seniority number. This novel partitioning scheme allows for a more compact description of CI-expansions \cite{Bytautas2011,Alcoba2013,Giesbertz2014}.

To show that the seniority-two sector can be decoupled from the seniority-zero-plus-two sectors, let $ \Psi_{\rm oo}^{(0)}$ be a seniority-zero wave function expansion with fully optimized one-particle functions and consider $\Psi_{\rm oo}^{(0,2)}$ to be a CI-expansion restricted to the seniority-zero-plus-two sectors (again assuming fully optimized one-particle functions), \emph{i.e.}, 
\begin{equation}
\Psi_{\rm oo}^{(0,2)} = \Psi_{\rm oo}^{(0)} +\Psi_{\rm oo}^{(2)}= (1 + \sum_{p \neq q} t_{pq} a^\dagger_p a_q  ) \Psi_{\rm oo}^{(0)}. 
\end{equation}
We are interested in the conditions for which $\Psi_{\rm oo}^{(0)}$ and $\Psi_{\rm oo}^{(0,2)}$ yield similar energy expectation values,
\begin{equation}\label{eq:decoupling-condition}
\langle \Psi_{\rm oo}^{(0,2)} |  \hat{H} | \Psi_{\rm oo}^{(0,2)} \rangle = E^{(0)},
\end{equation}
with $E^{(0)} = \langle \Psi_{\rm oo}^{(0)} |  \hat{H} | \Psi_{\rm oo}^{(0)} \rangle$ being the (ground state) energy expectation value of $ \Psi_{\rm oo}^{(0)}$.
Using the above identities, we can rewrite eq.~\eqref{eq:decoupling-condition} as
\begin{equation}\label{eq:decoupling}
\langle \Psi_{\rm oo}^{(0)} |  (1+ \hat{T}^\dagger ) \hat{H} (1+\hat{T}) | \Psi_{\rm oo}^{(0)} \rangle = E^{(0)},
\end{equation}
where we have introduced the excitation operator $T = \sum_{p\neq q} t_{pq} a^\dagger_p a_q$. The above equation is true to first order if $\hat{T} = {\bm \kappa}$, \emph{i.e.}, $(t_{pq})$ must be a skew-symmetric matrix with $\hat{T}^\dagger = -\hat{T}$. This requirement is equivalent to the invariance condition of the energy with respect to orbital rotations. Thus, if the seniority-zero-plus-two sectors can be written as a unitary transformation of the seniority-zero sector, we obtain similar results for the energy expectation value to first order. Then, it is sufficient to consider only the latter sector as a good approximation to the wave function. To decouple the seniority-two sector from the seniority-zero-plus-two sectors, we must have
\begin{equation}\label{eq:energy-diff}
\langle \Psi_{\rm oo}^{(0,2)} |  \hat{H} | \Psi_{\rm oo}^{(0,2)} \rangle  - \langle \Psi_{\rm oo}^{(0)} |  \hat{H} | \Psi_{\rm oo}^{(0)} \rangle = 0.
\end{equation}
This equation can be straightforwardly simplified. Considering only the first order terms in $\{t_{pq}\}$, we arrive at the decoupling condition
\begin{equation*}\label{eq:ps2-doscf-condition}
\langle \Psi_{\rm oo}^{(2)} |  \hat{H} | \Psi_{\rm oo}^{(0)} \rangle + c.c. = 0,
\end{equation*}
where $c.c.$ denotes the complex conjugate. This is a form of the generalized Brillouin theorem~\cite{Pian1981}. Substituting $\Psi_{\rm oo}^{(2)} = \sum_{p \neq q} t_{pq} a^\dagger_p a_q \Psi_{\rm oo}^{(0)} $ in the above equation, we can rewrite the first term as
\begin{equation}\label{eq:ps2-doscf}
\sum_{p \neq q} t_{pq} \langle  a_p^\dagger a_q  \Psi_{\rm oo}^{(0)} | \hat{H} | \Psi_{\rm oo}^{(0)} \rangle.
\end{equation}
In section~\ref{sec:ps2}, we will impose the stronger requirement that this term vanishes separately,
\begin{equation}\label{eq:ps2-doscf-final}
\langle  a_p^\dagger a_q  \Psi_{\rm oo}^{(0)} | \hat{H} | \Psi_{\rm oo}^{(0)} \rangle = 0 \quad \forall \, p \neq q,
\end{equation}
as would be the case for the Brillouin theorem in HF theory.

\section{Projected-Seniority-Two Orbital Optimization}\label{sec:ps2}
To derive the working equations for an orbital optimization scheme that aims at decoupling the seniority-two sector, we make two assumptions. First, we assume that ${\rm AP1roG}$ is a good approximation to ${\rm DOCI}$ (a seniority-zero wave function). The good performance of AP1roG in approximating the DOCI wave function has been justified by many numerical examples (see, for instance, refs.~\cite{Limacher_2013,Piotrus_Mol-Phys,OO-AP1roG,pawel_jpca-2014,p-CCD}). Then, the new set of orbitals can be chosen by requiring that the projection of the seniority-two sector on the AP1roG reference wave function is zero, and eq.~\eqref{eq:ps2-doscf-final} becomes
\begin{equation}\label{eq:fullps2}
\langle  a_p^\dagger a_q  {\rm AP1roG} | \hat{H} | {\rm AP1roG} \rangle = 0.
\end{equation}
Yet, the complexity of solving the above set of equations scales factorially with system size. Including all seniority-zero Slater determinants on the left hand side of eq.~\eqref{eq:fullps2} is, thus, computationally intractable. A pragmatic, but computationally feasible, approximation is to consider only singly pair-excited determinants with respect to $|\Phi_0 \rangle$. Therefore, our second assumption is to keep only the single-pair-excited determinants $\Phi_{i \bar{i}}^{a \bar{a}}$. Eq.~\eqref{eq:fullps2}, then, reduces to
\begin{equation}\label{eq:ps2}
\langle \Phi_0 + \sum_{i,a} c_i^a \Phi_{i \bar{i}}^{a \bar{a}} | a_q^\dagger a_p \big( e^{- \bm \kappa} \hat{H} e^{\bm \kappa} \big) |
        {\rm AP1roG} \rangle = 0,
\end{equation}
where $\bm \kappa$ is again the generator of orbital rotations and the Hamiltonian has been written explicitly in terms of the rotated basis. The $\{\kappa_{pq}\}$ are optimized such that eq.~\eqref{eq:ps2} is fulfilled. We will call eq.~\eqref{eq:ps2} the (approximate) projected-seniority-two (PS2) condition and the left hand side of eq.~\eqref{eq:ps2} the PS2 orbital gradient. Specifically, the PS2-orbital gradient is approximated for the case where $p > q$. Based on observations from ref.~\citenum{Bytautas2011}, the exact PS2 condition (eq.~\eqref{eq:fullps2}) will be valid for symmetry-adapted molecular orbitals, and thus the optimized orbitals will transform according to the irreducible representations of the molecular point group. Yet, this might not be true for the approximate PS2 condition (eq.~\eqref{eq:ps2}). However, we observed that the PS2 orbital optimization scheme preserves spatial symmetry in most cases. 
In this respect the PS2-AP1roG approach is similar to the Brueckner coupled cluster~\cite{Crawford2000} method. 
Note that the PS2 orbital gradient and the vOO orbital gradient differ and thus cannot be zero at the same time. 
The general exceptions are two-electron systems, where both the variational orbital optimization scheme and the projected-seniority-two approaches are equivalent to full-CI.

\begin{table*}
\caption{Spectroscopic constants: bond distances (R$_{\rm e}$), potential energy depths (D$_{\rm e}$) and harmonic vibrational frequencies ($\omega_{\rm e}$) for the ground state of C$_2$ and LiF (cc-pVDZ). Differences with respect to reference data are listed in parenthesis.
PS2-AP1roG: projected-seniority-two orbital optimized AP1roG; vOO-AP1roG(RHF)/(PS2): variational orbital optimization of AP1roG using RHF/PS2-optimized molecular orbitals as initial guess; vOO-AP1roG(BS): variational orbital optimization of AP1roG with symmetry-broken molecular orbitals.
}\label{tbl:diatomics}
{
\begin{tabular}{cllllll} 
\hline \hline
&
Method &
R$_{\rm e}$ [\AA]&&
D$_{\rm e}$ [eV] &&
$\omega_{\rm e}$ [cm$^{-1}$]\\
\hline
\multirow{8}{*}{{C$_2$ ($^1\Sigma^+_g$)}}
&PS2-AP1roG                      &   1.243 ($-$0.030) && 5.336 ($-$0.288) && 1949.9 ($+$136.3)\phantom{1}\\
&vOO-AP1roG(BS)                  &   1.271 ($-$0.002) && 5.359 ($-$0.265) && 1475.2 ($-$338.5)\\
&vOO-AP1roG(PS2)                 &   1.240 ($-$0.033) && 5.356 ($-$0.268) && 1939.6 ($+$125.9)\\
&vOO-AP1roG(RHF)\cite{pawel_jpca-2014}  &   1.240 ($-$0.033) && 5.414 ($-$0.210) && 1930.0 ($+$116.3) \\
&DMRG~\cite{CheMPS2}             &   1.269 ($-$0.004) && 5.628 ($+$0.004) && 1811.4 ($-$2.3)\phantom{11} \\
&FCI~\cite{Evangelista_2011}     &   1.273 ($+$0.000) && --               && 1812.9 ($-$0.8)\phantom{11} \\
\cline{2-7}
&C-MRCI+Q~\cite{Peterson1995}    & \multicolumn{1}{l}{1.273} && \multicolumn{1}{l}{5.624} &&\multicolumn{1}{l}{1813.7} \\
\hline
\multirow{6}{*}{{LiF ($^1\Sigma^-$)}}
&PS2-AP1roG         	         &   1.584 ($-$0.025) && 5.836 ($+$0.081) && 886.5 \\
&vOO-AP1roG(BS)                  &   1.586 ($-$0.023) && 4.583 ($-$1.172) && 868.0 \\
&vOO-AP1roG(PS2)                 &   1.580 ($-$0.029) && 4.567 ($-$1.188) && 913.4 \\
&DMRG$^{\rm a}$~\cite{LegezaLiF}          &   1.620 ($-$0.011) && 5.245 ($-$0.510) && 893.7 \\
\cline{2-7}
&MRCI(Q)-C3$_2$~\cite{Varandas2009}    & \multicolumn{1}{l}{1.609} && \multicolumn{1}{l}{5.755} &&\multicolumn{1}{l}{--} \\
\hline \hline
\end{tabular}
\begin{tablenotes}\scriptsize
\item $^{\rm a}$ cc-pVDZ basis set with additional diffuse functions on the F atom\\
\end{tablenotes}
}
\end{table*}

\section{Computational Details}
All vOO-AP1roG calculations have been performed with the \textsc{Horton} program package \cite{Horton13}, while the PS2 orbital optimization scheme for AP1roG has been implemented in a developer's version of the \textsc{Horton} program package. The variational and PS2 orbital optimizations were allowed to relax without any constraints on spatial symmetry, \emph{i.e.}, all calculations have been performed using C$_1$ point group symmetry. In our vOO-AP1roG calculations, two different initial guesses were applied: (i) PS2-AP1roG optimized orbitals, abbreviated by the acronym vOO-AP1roG(PS2), and (ii) restricted HF (RHF) orbitals. For PS2-AP1roG, only canonical RHF orbitals have been used as initial guess orbitals. 

 For the optimization of the molecular orbital basis (both vOO-AP1roG and PS2-AP1roG), we use a Newton--Raphson optimizer and a diagonal approximation of the orbital Hessian/Jacobian to obtain the rotated set of orbital expansion coefficients. In particular, the approximate Hessian contains terms at most linear in the geminal expansion coefficients and Lagrange multipliers.

As computational examples, we chose the dissociation process of the C$_2$ and LiF molecules, and the symmetric stretching of the CH$_2$ molecule and linear hydrogen chains including 18, 34 and 50 hydrogen atoms. The potential energy curves of all studied diatomic molecules were obtained by varying bond-lengths in the range of $1.4-8.0$~\AA~ for the LiF and $1.1-3.2$~\AA~for the C$_2$ molecule, respectively. The points on the resulting potential energy curve were used for a subsequent generalized Morse function~\cite{Coxon_1992} fit to obtain the equilibrium bond lengths (R$_{\rm e}$) and potential energy depths (D$_{\rm e}$). The harmonic vibrational frequencies ($\omega_{\rm e}$) were calculated numerically using the five-point finite difference stencil~\cite{Abramowitz}.

For the C$_2$ and LiF molecules we have employed a cc-pVDZ~\cite{dunning_b,emsl-basis-1,emsl-basis-2} basis set. 
The calculations on the CH$_2$ molecule were performed using the cc-pCVDZ~\cite{cc-pCVDZ} basis set, while for the hydrogen chain molecules the STO-6G~\cite{STO-6G-H-Ne} basis set was used to allow for a direct comparison with density matrix renormalization group (DMRG) reference data taken from ref.~\cite{Hachmann_H50}.

\section{Numerical Analysis}
For the C$_2$ molecule, two different solutions are obtained if canonical RHF orbitals are used: (i) broken-symmetry (BS) and (ii) symmetry-adapted molecular orbitals, labeled by vOO-AP1roG(BS) and vOO-AP1roG(RHF), respectively. By contrast, the PS2 scheme provides only a symmetry-adapted solution \footnote{We should note that all calculations are performed in C$_1$ symmetry (see computational details). Therefore, molecular orbitals cannot be labeled according to an irreducible representation of the molecular point group. The term 'symmetry-adapted' is used to emphasize that the optimized molecular orbitals are not symmetry-broken (\emph{e.g.}, localized, hybrid, etc.); this does not imply that the molecular orbitals transform as an irreducible representation of the molecular point group.}.

\begin{figure}[h]
\centering
\includegraphics[width=1.0\linewidth]{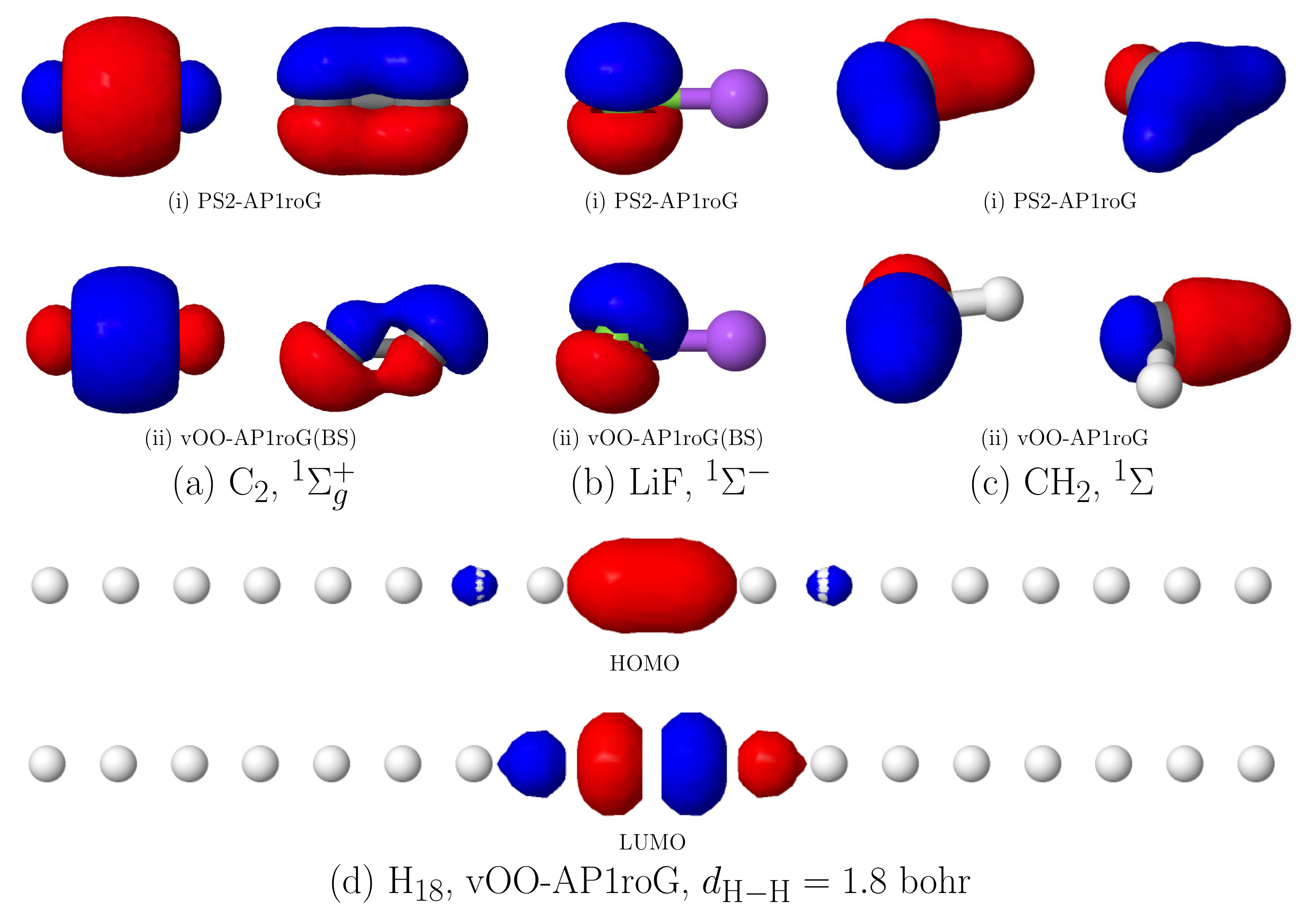}
\caption{Selected optimized valence orbitals for (a) C$_2$, (b) LiF, (c) CH$_2$ using different orbital optimization schemes and for (d) H$_{18}$ employing variational orbital optimization. For C$_2$, LiF, and CH$_2$, only the bonding orbitals are shown. The orbitals were visualized using the Jmol software package~\cite{Jmol}.
}
\label{fig:orbitals}
\end{figure}

Selected C$_2$ valence molecular orbitals are shown in Figure~\ref{fig:orbitals}(a). PS2-AP1roG, vOO-AP1roG(PS2) and vOO-AP1roG(RHF) yield similar solutions and thus only the PS2-AP1roG results are shown. For all internuclear distances, PS2-optimized orbitals closely resemble typical $\sigma$-, $\sigma^*$-, $\pi$- and $\pi^*$-orbitals, while the broken-symmetry ones can be identified as bonding and antibonding combinations of hybrid orbitals. We should note, that close to the dissociation limit ($r_{\rm C-C} \geq 2.0$ \AA), all orbital optimization schemes yield symmetry-adapted solutions.

The spectroscopic constants (equilibrium distance R$_{\rm e}$, potential energy depth D$_{\rm e}$ and harmonic vibrational frequency $\omega_{\rm e}$) of C$_2$ for all optimization schemes are presented in Table~\ref{tbl:diatomics}. In general, the spectroscopic constants obtained from PS2-AP1roG, vOO-AP1roG(PS2) and vOO-AP1roG(RHF) are similar and the deviations are within chemical accuracy (chemical accuracy corresponds to ca.~$0.04$ eV or 1.6 m$E_h$). Larger differences can be observed for the potential energy depth and amount to 0.078 eV (ca.~$2.9$ m$E_h$). As expected, $\omega_{\rm e}$ obtained from broken-symmetry solutions shows the largest deviations being considerably underestimated compared to MRCI reference data, though the equilibrium distance is in near perfect agreement with MRCI \cite{McLean1985}.

Selected optimized molecular orbitals for LiF are displayed in Figure~\ref{fig:orbitals}(b). Variational orbital optimization yields two distinct solutions: (i) broken-symmetry (vOO-AP1roG(BS)) and (ii) symmetry-adapted molecular orbitals (vOO-AP1roG(PS2)). Using a canonical RHF initial guess, we obtain only a broken-symmetry solution, while no symmetry-breaking is observed when PS2-AP1roG optimized orbitals are taken as initial guess. As observed for the C$_2$ molecule, vOO-AP1roG(BS) yields hybrid molecular orbitals (see Figure~\ref{fig:orbitals}(b-ii)), while spatial symmetry is retained in the PS2 scheme.

In general, equilibrium distances and harmonic vibrational frequencies are similar for all orbital optimization frameworks and initial guess orbitals, though differences in $\omega_{\rm e}$ are more pronounced and amount to ca.~$30$ cm$^{-1}$ (see Table~\ref{tbl:diatomics}). The potential energy depth is nearly independent on the initial guess if the orbitals are optimized variationally. Note, however, that D$_{\rm e}$ is underestimated by more than 1 eV. This can be explained by missing dynamic electron correlation effects around the equilibrium structure. Remarkably, the PS2-AP1roG scheme yields a potential energy depth that is in very good agreement with MRCI reference data. However, this might be coincidental because the PS2 orbital optimization (fulfillment of eq.~\eqref{eq:ps2}) is difficult to converge for stretched interatomic distances and local minima are easily encountered.

Our next example represents the CH$_2$ molecule. Table~\ref{tbl:ch2} summarizes energy differences between the equilibrium geometry $r_e$ and a symmetrically stretched geometry $2r_e$. In general, both the variational and PS2 orbital optimization schemes yield similar energy differences (deviations are smaller than 0.6 m$E_h$) that are in good agreement with FCI reference data (differences are ca.~$4$ m$E_h$). As expected, variational orbital optimization results in lower total energies than the PS2 approach (ca.~30 m$E_h$). Note, however, that the variational orbital optimization converges to the same stationary point regardless of the initial guess orbitals (RHF or PS2-optimized; see Table~\ref{tbl:ch2}).

Selected valence molecular orbitals are displayed in Figure~\ref{fig:orbitals}(c) for PS2-AP1roG (i) and vOO-AP1roG (ii) at equilibrium geometry. While variational orbital optimization results in localized, \emph{i.e.}, symmetry-broken, molecular orbitals, the PS2 optimization scheme yields delocalized molecular orbitals. We should note that for the symmetrically stretched geometry, both PS2-AP1roG and vOO-AP1roG provide localized solutions. These molecular orbitals are similar to those presented in Figure~\ref{fig:orbitals}(c-ii) and are thus not shown.

\begin{table}[t]
\caption{Energy difference for CH$_2$ between the equilibrium geometry ($r_e$) and symmetrically stretched geometry ($2r_e$) using a cc-pCVDZ basis set.
The equilibrium geometry and FCI reference data are taken from ref.~\cite{Kohn2005}.
PS2-AP1roG: projected-seniority-two orbital optimized AP1roG; vOO-AP1roG(BS)/(PS2): variational orbital optimization of AP1roG using RHF/PS2-optimized molecular orbitals as initial guess.
}\label{tbl:ch2}
{\scriptsize
\begin{tabular}{lcccccl} 
\hline \hline
Method
&& $E(1r_e)\,[E_{h}]$ 
&& $E(2r_e)\,[E_{h}]$
&& \multicolumn{1}{c}{$\Delta E\, [{\rm m}E_{h}]$}\\
\hline
RHF              && $-$38.881405  && $-$38.549362 && 332.043 ($-$109.753) \\
PS2-AP1roG       && $-$38.965382  && $-$38.739566 && 225.816 ($-$3.523) \\
vOO-AP1roG(BS)   && $-$38.995659  && $-$38.769222 && 226.436 ($-$4.147) \\
vOO-AP1roG(PS2)  && $-$38.995659  && $-$38.769222 && 226.436 ($-$4.147) \\\hline
FCI~\cite{Kohn2005}      && $-$39.061338  && $-$38.839048 && 222.290\\
\hline\hline
\end{tabular}
}
\end{table}

Finally, we discuss the performance of the PS2 optimization scheme on the symmetric dissociation of linear hydrogen chains containing up to 50 hydrogen atoms. This particular example is a commonly-used molecular model for strongly-correlated systems and remains a remarkably difficult task for conventional quantum-chemistry approaches~\cite{Hachmann_H50,Gustavo_H50,Stella_H50,DMFT_H50}.
Furthermore, studying the (symmetric) dissociation of hydrogen chains allows us to numerically assess to what extent the PS2 optimization procedure violates size-extensivity.
\begin{table}[t]
\caption{Total energy per hydrogen atom as a function of the interatomic distance $d_{\rm H-H}$ (in bohr) for hydrogen chains of different length in $E_h$.
PS2-AP1roG: projected-seniority-two orbital optimized AP1roG; vOO-AP1roG: variational orbital optimization of AP1roG.
}\label{tbl:hchains}
{\scriptsize
\begin{tabular}{lccccccc} 
\hline \hline
& \multicolumn{3}{c}{PS2-AP1roG}
&
& \multicolumn{3}{c}{vOO-AP1roG}
\\
$d_{\rm H-H}$
& H$_{18}$
& H$_{34}$
& H$_{50}$
&
& H$_{18}$
& H$_{34}$
& H$_{50}$
\\
\hline
1.00 &  -0.356124 & -0.343693 & -0.338924 &&-0.357157 & -0.344333 & -0.339627\\
1.20 &  -0.460062 & -0.452774 & -0.451430 &&-0.461106 & -0.454708 & -0.452395\\
1.40 &  -0.508042 & -0.504679 & -0.503575 &&-0.509569 & -0.506178 & -0.504960\\
1.60 &  -0.525773 & -0.525657 & -0.525241 &&-0.526120 & -0.527844 & -0.527170\\
1.80 &  -0.532758 & -0.531278 & -0.531427 &&-0.534744 & -0.533653 & -0.533261\\
2.00 &  -0.529623 & -0.528946 & -0.528819 &&-0.531648 & -0.530961 & -0.530714\\
2.40 &  -0.512675 & -0.512289 & -0.512507 &&-0.515489 & -0.515091 & -0.514947\\
2.80 &  -0.494904 & -0.494510 & -0.494592 &&-0.497940 & -0.497595 & -0.497470\\
3.20 &  -0.482201 & -0.481687 & -0.481661 &&-0.485099 & -0.484799 & -0.484691\\
3.60 &  -0.475189 & -0.474789 & -0.474715 &&-0.477703 & -0.477483 & -0.477404\\
4.20 &  -0.471527 & -0.471323 & -0.471263 &&-0.473057 & -0.472954 & -0.472917\\
\hline\hline
\end{tabular}
}
\end{table}

Table~\ref{tbl:hchains} summarizes the total energy per hydrogen atom as a function of the internuclear separation for chains containing 18, 34, and 50 hydrogen atoms. Only one solution is obtained in the variational orbital optimization irrespective of the initial guess used, \emph{i.e.}, canonical RHF or PS2-optimized orbitals, and is labeled by the acronym vOO-AP1roG. In particular, both orbital-optimization protocols result in symmetry-broken, localized molecular orbitals for all chain lengths and interatomic distances. As an example, the highest-occupied and lowest-unoccupied molecular orbitals (HOMO and LUMO) with respect to the HF reference determinant are shown in Figure~\ref{fig:orbitals}(d).

\begin{figure}[h]
\centering
\includegraphics[width=1.0\linewidth]{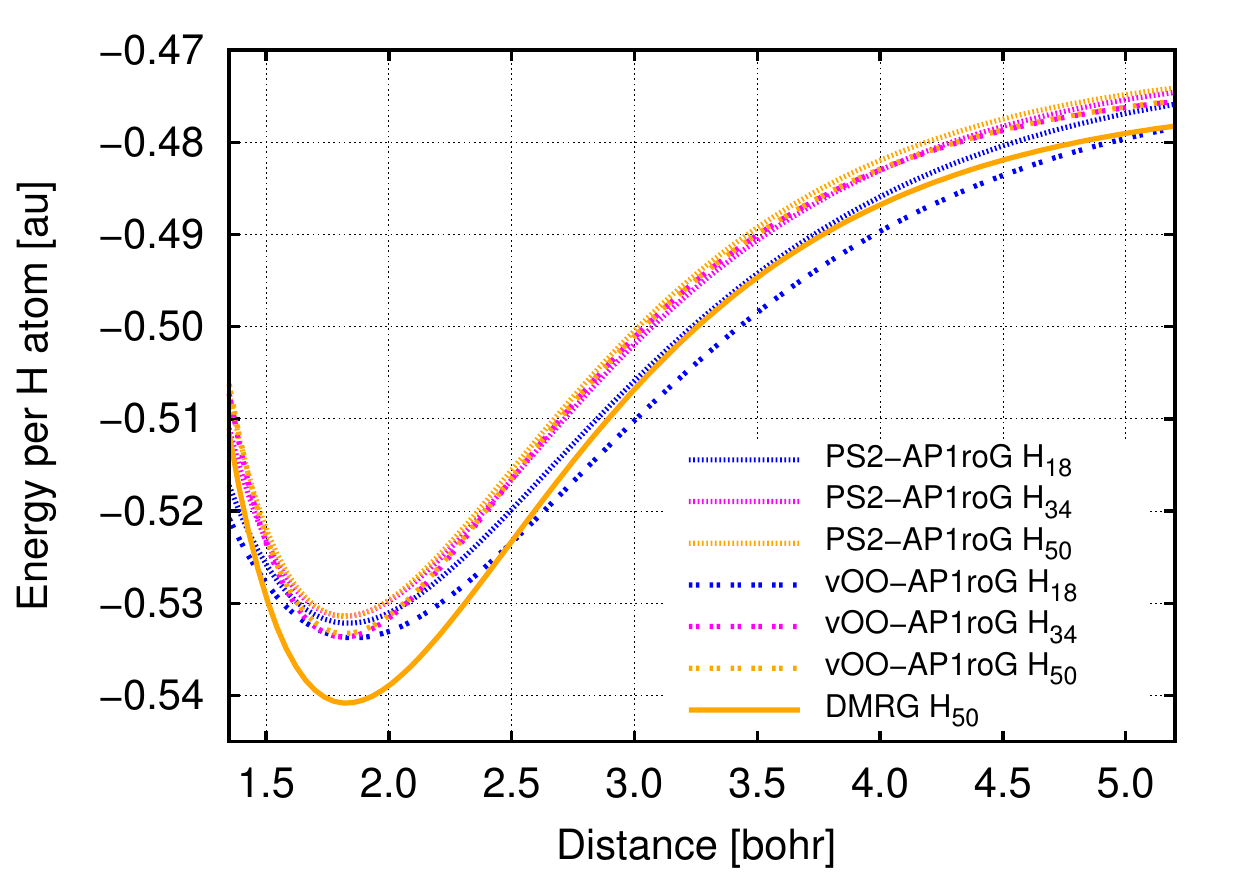}
\caption{Fitted total energy per hydrogen atom curves for hydrogen chains of different lengths using the PS2 and variational orbital optimization procedures. The DMRG reference data has been taken from ref.~\cite{Hachmann_H50}.
}
\label{fig:chains}
\end{figure}

For both orbital optimization schemes, the total energy per hydrogen atom varies marginally if the number of hydrogen atoms in the chains is increased (see Table~\ref{tbl:hchains}). As expected, the variational orbital optimization yields lower total ground state energies (and thus lower values for the energy per atom) than the PS2 scheme. A generalized Morse potential fit to the energy-per-atom data is displayed in Figure~\ref{fig:chains} for both orbital optimization schemes and different chain lengths. Specifically, close to the equilibrium distance (ca.~1.8 bohr) and in the vicinity of dissociation, the variations in the total energy per atom become considerably smaller than chemical accuracy. Larger differences can be observed for squeezed chains and for the smallest chain length studied in this work. We should emphasize that both orbital optimization procedures yield essentially parallel energy-per-atom curves (see Figure~\ref{fig:chains}) with differences only up to 3 m$E_h$. Our numerical analysis, thus, suggests that size-extensivity is not affected by the PS2 orbital optimization scheme.

Finally, Figure~\ref{fig:chains} shows the fitted DMRG reference energy-per-atom curve for the symmetric stretching problem of H$_{50}$. The overall agreement of PS2-/vOO-AP1roG with DMRG is good~\cite{OO-AP1roG}. Larger deviations from the DMRG reference are found around the equilibrium distance and originate from dynamic correlation effects that cannot be captured by AP1roG~\cite{OO-AP1roG,Johnson_2013,Limacher_2013}. For stretched internuclear distances, the PS2-/vOO-AP1roG energy-per-atom curves are closely parallel to the DMRG reference.

\section{Conclusions}
We have presented an alternative orbital optimization scheme for the AP1roG wave function that does not rely on the variational principle. Our approach is motivated by the observation that an orbital-optimized CI expansion constructed from the seniority-zero sector gives similar results to an orbital-optimized CI expansion constructed from the seniority-zero-plus-two sectors~\cite{Bytautas2011}. We have compared this new orbital-optimization scheme to its variational counterpart for the C$_2$, LiF, CH$_2$ and linear hydrogen chain molecules containing 18, 34 and 50 hydrogen atoms.

In general, both orbital optimization procedures yield similar spectroscopic constants and energy differences. While the variational orbital optimization favors localized (symmetry-broken) solutions, the PS2 approach preferably retains spatial symmetry and yields delocalized molecular orbitals. However, PS2 orbital optimization results in localized orbitals if strong correlation becomes important, like, for instance, in stretched bonds or linear hydrogen chains. Furthermore, a numerical analysis on linear hydrogen chain molecules of different length suggests that PS2 does not violate size-extensivity.

In practical applications, we recommend the variational orbital optimization. In contrast to the PS2 scheme that suffers from convergence difficulties for molecules with stretched bonds, the variational orbital optimization seems to be more robust and less sensitive to the initial guess. However, an alternating variational-PS2-orbital-optimization procedure can be applied to increase convergence, in general, and to dodge local minima.

Different, less stringent approximations to the PS2-orbital gradient are currently being investigated in our laboratory.

\section{Acknowledgment}
We gratefully acknowledge financial support from the Natural Sciences and Engineering Research Council of Canada. K.B. acknowledges the financial support from the Swiss National Science Foundation (P2EZP2 148650). P.A.J. acknowledges funding from a Vanier Canada Graduate Scholarship. 
P.B. with S.D.B. and D.V.N acknowledge financial support from FWO-Flanders and the Research Council of Ghent University. S.D.B is an FWO postdoctoral fellow.

We thank Gustavo E.~Scuseria and Tamar~Stein for many helpful discussions.

\bibliography{rsc}

\begin{thebibliography}{67}%
\makeatletter
\providecommand \@ifxundefined [1]{%
 \@ifx{#1\undefined}
}%
\providecommand \@ifnum [1]{%
 \ifnum #1\expandafter \@firstoftwo
 \else \expandafter \@secondoftwo
 \fi
}%
\providecommand \@ifx [1]{%
 \ifx #1\expandafter \@firstoftwo
 \else \expandafter \@secondoftwo
 \fi
}%
\providecommand \natexlab [1]{#1}%
\providecommand \enquote  [1]{``#1''}%
\providecommand \bibnamefont  [1]{#1}%
\providecommand \bibfnamefont [1]{#1}%
\providecommand \citenamefont [1]{#1}%
\providecommand \href@noop [0]{\@secondoftwo}%
\providecommand \href [0]{\begingroup \@sanitize@url \@href}%
\providecommand \@href[1]{\@@startlink{#1}\@@href}%
\providecommand \@@href[1]{\endgroup#1\@@endlink}%
\providecommand \@sanitize@url [0]{\catcode `\\12\catcode `\$12\catcode
  `\&12\catcode `\#12\catcode `\^12\catcode `\_12\catcode `\%12\relax}%
\providecommand \@@startlink[1]{}%
\providecommand \@@endlink[0]{}%
\providecommand \url  [0]{\begingroup\@sanitize@url \@url }%
\providecommand \@url [1]{\endgroup\@href {#1}{\urlprefix }}%
\providecommand \urlprefix  [0]{URL }%
\providecommand \Eprint [0]{\href }%
\providecommand \doibase [0]{http://dx.doi.org/}%
\providecommand \selectlanguage [0]{\@gobble}%
\providecommand \bibinfo  [0]{\@secondoftwo}%
\providecommand \bibfield  [0]{\@secondoftwo}%
\providecommand \translation [1]{[#1]}%
\providecommand \BibitemOpen [0]{}%
\providecommand \bibitemStop [0]{}%
\providecommand \bibitemNoStop [0]{.\EOS\space}%
\providecommand \EOS [0]{\spacefactor3000\relax}%
\providecommand \BibitemShut  [1]{\csname bibitem#1\endcsname}%
\let\auto@bib@innerbib\@empty
\bibitem [{\citenamefont {Hurley}\ \emph {et~al.}(1953)\citenamefont {Hurley},
  \citenamefont {Lennard-Jones},\ and\ \citenamefont {Pople}}]{Hurley_1953}%
  \BibitemOpen
  \bibfield  {author} {\bibinfo {author} {\bibfnamefont {A.~C.}\ \bibnamefont
  {Hurley}}, \bibinfo {author} {\bibfnamefont {J.}~\bibnamefont
  {Lennard-Jones}}, \ and\ \bibinfo {author} {\bibfnamefont {J.~A.}\
  \bibnamefont {Pople}},\ }\href@noop {} {\bibfield  {journal} {\bibinfo
  {journal} {Proc. R. Soc. Lond. A}\ }\textbf {\bibinfo {volume} {220}},\
  \bibinfo {pages} {446} (\bibinfo {year} {1953})}\BibitemShut {NoStop}%
\bibitem [{\citenamefont {Kutzelnigg}(1964)}]{Kutzelnigg_1964}%
  \BibitemOpen
  \bibfield  {author} {\bibinfo {author} {\bibfnamefont {W.}~\bibnamefont
  {Kutzelnigg}},\ }\href@noop {} {\bibfield  {journal} {\bibinfo  {journal} {J.
  Chem. Phys.}\ }\textbf {\bibinfo {volume} {40}},\ \bibinfo {pages} {3640}
  (\bibinfo {year} {1964})}\BibitemShut {NoStop}%
\bibitem [{\citenamefont {Coleman}(1965)}]{Coleman_1965}%
  \BibitemOpen
  \bibfield  {author} {\bibinfo {author} {\bibfnamefont {A.~J.}\ \bibnamefont
  {Coleman}},\ }\href@noop {} {\bibfield  {journal} {\bibinfo  {journal} {J.
  Math. Phys.}\ }\textbf {\bibinfo {volume} {6}},\ \bibinfo {pages} {1425}
  (\bibinfo {year} {1965})}\BibitemShut {NoStop}%
\bibitem [{\citenamefont {Kutzelnigg}(1965)}]{Kutzelnigg_1965}%
  \BibitemOpen
  \bibfield  {author} {\bibinfo {author} {\bibfnamefont {W.}~\bibnamefont
  {Kutzelnigg}},\ }\href@noop {} {\bibfield  {journal} {\bibinfo  {journal}
  {Theoret. Chim. Acta}\ }\textbf {\bibinfo {volume} {3}},\ \bibinfo {pages}
  {241} (\bibinfo {year} {1965})}\BibitemShut {NoStop}%
\bibitem [{\citenamefont {Miller}\ and\ \citenamefont
  {Klaus}(1968)}]{Miller1968}%
  \BibitemOpen
  \bibfield  {author} {\bibinfo {author} {\bibfnamefont {K.~J.}\ \bibnamefont
  {Miller}}\ and\ \bibinfo {author} {\bibfnamefont {R.}~\bibnamefont {Klaus}},\
  }\href@noop {} {\bibfield  {journal} {\bibinfo  {journal} {J. Chem. Phys.}\
  }\textbf {\bibinfo {volume} {48}},\ \bibinfo {pages} {3444} (\bibinfo {year}
  {1968})}\BibitemShut {NoStop}%
\bibitem [{\citenamefont {Ortiz}\ \emph {et~al.}(1981)\citenamefont {Ortiz},
  \citenamefont {Weiner},\ and\ \citenamefont {Ohrn}}]{Ortiz_1981}%
  \BibitemOpen
  \bibfield  {author} {\bibinfo {author} {\bibfnamefont {J.~V.}\ \bibnamefont
  {Ortiz}}, \bibinfo {author} {\bibfnamefont {B.}~\bibnamefont {Weiner}}, \
  and\ \bibinfo {author} {\bibfnamefont {Y.}~\bibnamefont {Ohrn}},\ }\href@noop
  {} {\bibfield  {journal} {\bibinfo  {journal} {Int. J. Quantum Chem.}\
  }\textbf {\bibinfo {volume} {S15}},\ \bibinfo {pages} {113} (\bibinfo {year}
  {1981})}\BibitemShut {NoStop}%
\bibitem [{\citenamefont {Surj\'{a}n}(1999)}]{Surjan_1999}%
  \BibitemOpen
  \bibfield  {author} {\bibinfo {author} {\bibfnamefont {P.~R.}\ \bibnamefont
  {Surj\'{a}n}},\ }in\ \href@noop {} {\emph {\bibinfo {booktitle} {Correlation
  and Localization}}}\ (\bibinfo  {publisher} {Springer},\ \bibinfo {year}
  {1999})\ pp.\ \bibinfo {pages} {63--88}\BibitemShut {NoStop}%
\bibitem [{\citenamefont {Surj\'{a}n}\ \emph {et~al.}(2012)\citenamefont
  {Surj\'{a}n}, \citenamefont {Szabados}, \citenamefont {Jeszenszki},\ and\
  \citenamefont {Zoboki}}]{Surjan_2012}%
  \BibitemOpen
  \bibfield  {author} {\bibinfo {author} {\bibfnamefont {P.~R.}\ \bibnamefont
  {Surj\'{a}n}}, \bibinfo {author} {\bibfnamefont {A.}~\bibnamefont
  {Szabados}}, \bibinfo {author} {\bibfnamefont {P.}~\bibnamefont
  {Jeszenszki}}, \ and\ \bibinfo {author} {\bibfnamefont {T.}~\bibnamefont
  {Zoboki}},\ }\href@noop {} {\bibfield  {journal} {\bibinfo  {journal} {J.
  Math. Chem.}\ }\textbf {\bibinfo {volume} {50}},\ \bibinfo {pages} {534}
  (\bibinfo {year} {2012})}\BibitemShut {NoStop}%
\bibitem [{\citenamefont {Kutzelnigg}(2012)}]{Kutzelnigg2012}%
  \BibitemOpen
  \bibfield  {author} {\bibinfo {author} {\bibfnamefont {W.}~\bibnamefont
  {Kutzelnigg}},\ }\href@noop {} {\bibfield  {journal} {\bibinfo  {journal}
  {Chem. Phys.}\ }\textbf {\bibinfo {volume} {401}},\ \bibinfo {pages} {119}
  (\bibinfo {year} {2012})}\BibitemShut {NoStop}%
\bibitem [{\citenamefont {McWeeny}\ and\ \citenamefont
  {Sutcliffe}(1963)}]{McWeeny1963}%
  \BibitemOpen
  \bibfield  {author} {\bibinfo {author} {\bibfnamefont {R.}~\bibnamefont
  {McWeeny}}\ and\ \bibinfo {author} {\bibfnamefont {B.~T.}\ \bibnamefont
  {Sutcliffe}},\ }\href@noop {} {\bibfield  {journal} {\bibinfo  {journal}
  {Proc. R. Soc. Lond. A}\ }\textbf {\bibinfo {volume} {273}},\ \bibinfo
  {pages} {103} (\bibinfo {year} {1963})}\BibitemShut {NoStop}%
\bibitem [{\citenamefont {Cassam-Chena\"{\i}}(2006)}]{Cassam-Chenai2006}%
  \BibitemOpen
  \bibfield  {author} {\bibinfo {author} {\bibfnamefont {P.}~\bibnamefont
  {Cassam-Chena\"{\i}}},\ }\href@noop {} {\bibfield  {journal} {\bibinfo
  {journal} {J. Chem. Phys.}\ }\textbf {\bibinfo {volume} {124}},\ \bibinfo
  {pages} {194109} (\bibinfo {year} {2006})}\BibitemShut {NoStop}%
\bibitem [{\citenamefont {Cassam-Chena\"{\i}}\ and\ \citenamefont
  {Granucci}(2007)}]{Cassam-Chenai2007}%
  \BibitemOpen
  \bibfield  {author} {\bibinfo {author} {\bibfnamefont {P.}~\bibnamefont
  {Cassam-Chena\"{\i}}}\ and\ \bibinfo {author} {\bibfnamefont
  {G.}~\bibnamefont {Granucci}},\ }\href@noop {} {\bibfield  {journal}
  {\bibinfo  {journal} {Chem. Phys. Lett.}\ }\textbf {\bibinfo {volume}
  {450}},\ \bibinfo {pages} {151} (\bibinfo {year} {2007})}\BibitemShut
  {NoStop}%
\bibitem [{\citenamefont {Scuseria}\ and\ \citenamefont
  {Tsuchimochi}(2009)}]{scuseria2009constrained}%
  \BibitemOpen
  \bibfield  {author} {\bibinfo {author} {\bibfnamefont {G.~E.}\ \bibnamefont
  {Scuseria}}\ and\ \bibinfo {author} {\bibfnamefont {T.}~\bibnamefont
  {Tsuchimochi}},\ }\href@noop {} {\bibfield  {journal} {\bibinfo  {journal}
  {J. Chem. Phys.}\ }\textbf {\bibinfo {volume} {131}},\ \bibinfo {pages}
  {164119} (\bibinfo {year} {2009})}\BibitemShut {NoStop}%
\bibitem [{\citenamefont {Scuseria}\ \emph {et~al.}(2011)\citenamefont
  {Scuseria}, \citenamefont {Jim{\'e}nez-Hoyos}, \citenamefont {Henderson},
  \citenamefont {Samanta},\ and\ \citenamefont
  {Ellis}}]{scuseria2011projected}%
  \BibitemOpen
  \bibfield  {author} {\bibinfo {author} {\bibfnamefont {G.~E.}\ \bibnamefont
  {Scuseria}}, \bibinfo {author} {\bibfnamefont {C.~A.}\ \bibnamefont
  {Jim{\'e}nez-Hoyos}}, \bibinfo {author} {\bibfnamefont {T.~M.}\ \bibnamefont
  {Henderson}}, \bibinfo {author} {\bibfnamefont {K.}~\bibnamefont {Samanta}},
  \ and\ \bibinfo {author} {\bibfnamefont {J.~K.}\ \bibnamefont {Ellis}},\
  }\href@noop {} {\bibfield  {journal} {\bibinfo  {journal} {J. Chem. Phys.}\
  }\textbf {\bibinfo {volume} {135}},\ \bibinfo {pages} {124108} (\bibinfo
  {year} {2011})}\BibitemShut {NoStop}%
\bibitem [{\citenamefont {Limacher}\ \emph {et~al.}(2013)\citenamefont
  {Limacher}, \citenamefont {Ayers}, \citenamefont {Johnson}, \citenamefont
  {De~Baerdemacker}, \citenamefont {Van~Neck},\ and\ \citenamefont
  {Bultinck}}]{Limacher_2013}%
  \BibitemOpen
  \bibfield  {author} {\bibinfo {author} {\bibfnamefont {P.~A.}\ \bibnamefont
  {Limacher}}, \bibinfo {author} {\bibfnamefont {P.~W.}\ \bibnamefont {Ayers}},
  \bibinfo {author} {\bibfnamefont {P.~A.}\ \bibnamefont {Johnson}}, \bibinfo
  {author} {\bibfnamefont {S.}~\bibnamefont {De~Baerdemacker}}, \bibinfo
  {author} {\bibfnamefont {D.}~\bibnamefont {Van~Neck}}, \ and\ \bibinfo
  {author} {\bibfnamefont {P.}~\bibnamefont {Bultinck}},\ }\href@noop {}
  {\bibfield  {journal} {\bibinfo  {journal} {J. Chem. Theory Comput.}\
  }\textbf {\bibinfo {volume} {9}},\ \bibinfo {pages} {1394} (\bibinfo {year}
  {2013})}\BibitemShut {NoStop}%
\bibitem [{\citenamefont {Johnson}\ \emph {et~al.}(2013)\citenamefont
  {Johnson}, \citenamefont {Ayers}, \citenamefont {Limacher}, \citenamefont
  {De~Baerdemacker}, \citenamefont {Van~Neck},\ and\ \citenamefont
  {Bultinck}}]{Johnson_2013}%
  \BibitemOpen
  \bibfield  {author} {\bibinfo {author} {\bibfnamefont {P.~A.}\ \bibnamefont
  {Johnson}}, \bibinfo {author} {\bibfnamefont {P.~W.}\ \bibnamefont {Ayers}},
  \bibinfo {author} {\bibfnamefont {P.~A.}\ \bibnamefont {Limacher}}, \bibinfo
  {author} {\bibfnamefont {S.}~\bibnamefont {De~Baerdemacker}}, \bibinfo
  {author} {\bibfnamefont {D.}~\bibnamefont {Van~Neck}}, \ and\ \bibinfo
  {author} {\bibfnamefont {P.}~\bibnamefont {Bultinck}},\ }\href@noop {}
  {\bibfield  {journal} {\bibinfo  {journal} {Comput. Chem. Theory}\ }\textbf
  {\bibinfo {volume} {1003}},\ \bibinfo {pages} {101} (\bibinfo {year}
  {2013})}\BibitemShut {NoStop}%
\bibitem [{\citenamefont {Ellis}\ \emph {et~al.}(2013)\citenamefont {Ellis},
  \citenamefont {Martin},\ and\ \citenamefont {Scuseria}}]{ellis2013pair}%
  \BibitemOpen
  \bibfield  {author} {\bibinfo {author} {\bibfnamefont {J.~K.}\ \bibnamefont
  {Ellis}}, \bibinfo {author} {\bibfnamefont {R.~L.}\ \bibnamefont {Martin}}, \
  and\ \bibinfo {author} {\bibfnamefont {G.~E.}\ \bibnamefont {Scuseria}},\
  }\href@noop {} {\bibfield  {journal} {\bibinfo  {journal} {J. Chem. Theory
  Comput.}\ }\textbf {\bibinfo {volume} {9}},\ \bibinfo {pages} {2857}
  (\bibinfo {year} {2013})}\BibitemShut {NoStop}%
\bibitem [{\citenamefont {Tecmer}\ \emph {et~al.}(2014)\citenamefont {Tecmer},
  \citenamefont {Boguslawski}, \citenamefont {Limacher}, \citenamefont
  {Johnson}, \citenamefont {Chan}, \citenamefont {Verstraelen},\ and\
  \citenamefont {Ayers}}]{pawel_jpca-2014}%
  \BibitemOpen
  \bibfield  {author} {\bibinfo {author} {\bibfnamefont {P.}~\bibnamefont
  {Tecmer}}, \bibinfo {author} {\bibfnamefont {K.}~\bibnamefont {Boguslawski}},
  \bibinfo {author} {\bibfnamefont {P.~A.}\ \bibnamefont {Limacher}}, \bibinfo
  {author} {\bibfnamefont {P.~A.}\ \bibnamefont {Johnson}}, \bibinfo {author}
  {\bibfnamefont {M.}~\bibnamefont {Chan}}, \bibinfo {author} {\bibfnamefont
  {T.}~\bibnamefont {Verstraelen}}, \ and\ \bibinfo {author} {\bibfnamefont
  {P.~W.}\ \bibnamefont {Ayers}},\ }\href@noop {} {\bibfield  {journal}
  {\bibinfo  {journal} {J. Phys. Chem. A}\ }\textbf {\bibinfo {volume}
  {submitted}},\ \bibinfo {pages} {XX} (\bibinfo {year} {2014})}\BibitemShut
  {NoStop}%
\bibitem [{\citenamefont {Henderson}\ \emph {et~al.}(2014)\citenamefont
  {Henderson}, \citenamefont {Dukelsky}, \citenamefont {Scuseria},
  \citenamefont {Signoracci},\ and\ \citenamefont {Duguet}}]{p-CCD}%
  \BibitemOpen
  \bibfield  {author} {\bibinfo {author} {\bibfnamefont {T.~M.}\ \bibnamefont
  {Henderson}}, \bibinfo {author} {\bibfnamefont {J.}~\bibnamefont {Dukelsky}},
  \bibinfo {author} {\bibfnamefont {G.~E.}\ \bibnamefont {Scuseria}}, \bibinfo
  {author} {\bibfnamefont {A.}~\bibnamefont {Signoracci}}, \ and\ \bibinfo
  {author} {\bibfnamefont {T.}~\bibnamefont {Duguet}},\ }\href@noop {}
  {\bibfield  {journal} {\bibinfo  {journal} {arXiv preprint arXiv:1403.6818}\
  } (\bibinfo {year} {2014})}\BibitemShut {NoStop}%
\bibitem [{\citenamefont {Stein}\ \emph {et~al.}(2014)\citenamefont {Stein},
  \citenamefont {Henderson},\ and\ \citenamefont {Scuseria}}]{Tamar-p-CCD}%
  \BibitemOpen
  \bibfield  {author} {\bibinfo {author} {\bibfnamefont {T.}~\bibnamefont
  {Stein}}, \bibinfo {author} {\bibfnamefont {T.~M.}\ \bibnamefont
  {Henderson}}, \ and\ \bibinfo {author} {\bibfnamefont {G.~E.}\ \bibnamefont
  {Scuseria}},\ }\href@noop {} {\bibfield  {journal} {\bibinfo  {journal} {J.
  Chem. Phys. Comm.}\ }\textbf {\bibinfo {volume} {submitted}},\ \bibinfo
  {pages} {XX} (\bibinfo {year} {2014})}\BibitemShut {NoStop}%
\bibitem [{\citenamefont {Bratoz}\ and\ \citenamefont
  {Durand}(1965)}]{Bratoz1965}%
  \BibitemOpen
  \bibfield  {author} {\bibinfo {author} {\bibfnamefont {S.}~\bibnamefont
  {Bratoz}}\ and\ \bibinfo {author} {\bibfnamefont {P.}~\bibnamefont
  {Durand}},\ }\href@noop {} {\bibfield  {journal} {\bibinfo  {journal} {J.
  Chem. Phys.}\ }\textbf {\bibinfo {volume} {43}},\ \bibinfo {pages} {2670}
  (\bibinfo {year} {1965})}\BibitemShut {NoStop}%
\bibitem [{\citenamefont {Silver}(1969)}]{Silver_1969}%
  \BibitemOpen
  \bibfield  {author} {\bibinfo {author} {\bibfnamefont {D.~M.}\ \bibnamefont
  {Silver}},\ }\href@noop {} {\bibfield  {journal} {\bibinfo  {journal} {J.
  Chem. Phys.}\ }\textbf {\bibinfo {volume} {50}},\ \bibinfo {pages} {5108}
  (\bibinfo {year} {1969})}\BibitemShut {NoStop}%
\bibitem [{\citenamefont {Silver}(1970)}]{Silver_1970}%
  \BibitemOpen
  \bibfield  {author} {\bibinfo {author} {\bibfnamefont {D.~M.}\ \bibnamefont
  {Silver}},\ }\href@noop {} {\bibfield  {journal} {\bibinfo  {journal} {J.
  Chem. Phys.}\ }\textbf {\bibinfo {volume} {52}},\ \bibinfo {pages} {299}
  (\bibinfo {year} {1970})}\BibitemShut {NoStop}%
\bibitem [{\citenamefont {N\'{a}ray-Szab\'{o}}(1973)}]{APIG-1}%
  \BibitemOpen
  \bibfield  {author} {\bibinfo {author} {\bibfnamefont {G.}~\bibnamefont
  {N\'{a}ray-Szab\'{o}}},\ }\href@noop {} {\bibfield  {journal} {\bibinfo
  {journal} {J. Chem. Phys.}\ }\textbf {\bibinfo {volume} {58}},\ \bibinfo
  {pages} {1775} (\bibinfo {year} {1973})}\BibitemShut {NoStop}%
\bibitem [{\citenamefont {N\'{a}ray-Szab\'{o}}(1975)}]{APIG-2}%
  \BibitemOpen
  \bibfield  {author} {\bibinfo {author} {\bibfnamefont {G.}~\bibnamefont
  {N\'{a}ray-Szab\'{o}}},\ }\href@noop {} {\bibfield  {journal} {\bibinfo
  {journal} {Int. J. Qunatum Chem.}\ }\textbf {\bibinfo {volume} {9}},\
  \bibinfo {pages} {9} (\bibinfo {year} {1975})}\BibitemShut {NoStop}%
\bibitem [{\citenamefont {Surj\'{a}n}(1984)}]{Surjan-bond-1984}%
  \BibitemOpen
  \bibfield  {author} {\bibinfo {author} {\bibfnamefont {P.~R.}\ \bibnamefont
  {Surj\'{a}n}},\ }\href@noop {} {\bibfield  {journal} {\bibinfo  {journal}
  {Phys. Rev. A}\ }\textbf {\bibinfo {volume} {30}},\ \bibinfo {pages} {43}
  (\bibinfo {year} {1984})}\BibitemShut {NoStop}%
\bibitem [{\citenamefont {Surj\'{a}n}(1985)}]{Surjan-bond-1985}%
  \BibitemOpen
  \bibfield  {author} {\bibinfo {author} {\bibfnamefont {P.~R.}\ \bibnamefont
  {Surj\'{a}n}},\ }\href@noop {} {\bibfield  {journal} {\bibinfo  {journal}
  {Phys. Rev. A}\ }\textbf {\bibinfo {volume} {32}},\ \bibinfo {pages} {748}
  (\bibinfo {year} {1985})}\BibitemShut {NoStop}%
\bibitem [{\citenamefont {Surj\'{a}n}(1994)}]{Surjan-bond-1994}%
  \BibitemOpen
  \bibfield  {author} {\bibinfo {author} {\bibfnamefont {P.~R.}\ \bibnamefont
  {Surj\'{a}n}},\ }\href@noop {} {\bibfield  {journal} {\bibinfo  {journal}
  {Int. J. Quantum Chem.}\ }\textbf {\bibinfo {volume} {52}},\ \bibinfo {pages}
  {563} (\bibinfo {year} {1994})}\BibitemShut {NoStop}%
\bibitem [{\citenamefont {Surj\'{a}n}(1995)}]{Surjan-bond-1995}%
  \BibitemOpen
  \bibfield  {author} {\bibinfo {author} {\bibfnamefont {P.~R.}\ \bibnamefont
  {Surj\'{a}n}},\ }\href@noop {} {\bibfield  {journal} {\bibinfo  {journal}
  {Int. J. Quantum Chem.}\ }\textbf {\bibinfo {volume} {55}},\ \bibinfo {pages}
  {109} (\bibinfo {year} {1995})}\BibitemShut {NoStop}%
\bibitem [{\citenamefont {Rosta}\ and\ \citenamefont
  {Surj\'{a}n}(2000)}]{Surjan-bond-2000}%
  \BibitemOpen
  \bibfield  {author} {\bibinfo {author} {\bibfnamefont {E.}~\bibnamefont
  {Rosta}}\ and\ \bibinfo {author} {\bibfnamefont {P.~R.}\ \bibnamefont
  {Surj\'{a}n}},\ }\href
  {http://doi.wiley.com/10.1002/1097-461X\%282000\%2980\%3A2\%3C96\%3A\%3AAID-QUA4\%3E3.0.CO\%3B2-8}
  {\bibfield  {journal} {\bibinfo  {journal} {Int. J. Quantum Chem.}\ }\textbf
  {\bibinfo {volume} {80}},\ \bibinfo {pages} {96} (\bibinfo {year}
  {2000})}\BibitemShut {NoStop}%
\bibitem [{\citenamefont {Rosta}\ and\ \citenamefont
  {Surj\'{a}n}(2002)}]{Rosta2002}%
  \BibitemOpen
  \bibfield  {author} {\bibinfo {author} {\bibfnamefont {E.}~\bibnamefont
  {Rosta}}\ and\ \bibinfo {author} {\bibfnamefont {P.~R.}\ \bibnamefont
  {Surj\'{a}n}},\ }\href@noop {} {\bibfield  {journal} {\bibinfo  {journal} {J.
  Chem. Phys.}\ }\textbf {\bibinfo {volume} {116}},\ \bibinfo {pages} {878}
  (\bibinfo {year} {2002})}\BibitemShut {NoStop}%
\bibitem [{\citenamefont {Weinhold}\ and\ \citenamefont {Wilson}(1967)}]{DOCI}%
  \BibitemOpen
  \bibfield  {author} {\bibinfo {author} {\bibfnamefont {F.}~\bibnamefont
  {Weinhold}}\ and\ \bibinfo {author} {\bibfnamefont {E.~B.}\ \bibnamefont
  {Wilson}},\ }\href@noop {} {\bibfield  {journal} {\bibinfo  {journal} {J.
  Chem. Phys.}\ }\textbf {\bibinfo {volume} {46}},\ \bibinfo {pages} {2752}
  (\bibinfo {year} {1967})}\BibitemShut {NoStop}%
\bibitem [{\citenamefont {Parks}\ and\ \citenamefont
  {Parr}(1958)}]{Parks_1958}%
  \BibitemOpen
  \bibfield  {author} {\bibinfo {author} {\bibfnamefont {J.~M.}\ \bibnamefont
  {Parks}}\ and\ \bibinfo {author} {\bibfnamefont {R.~G.}\ \bibnamefont
  {Parr}},\ }\href@noop {} {\bibfield  {journal} {\bibinfo  {journal} {J. Chem.
  Phys.}\ }\textbf {\bibinfo {volume} {28}},\ \bibinfo {pages} {335} (\bibinfo
  {year} {1958})}\BibitemShut {NoStop}%
\bibitem [{\citenamefont {{Goddard III}}\ and\ \citenamefont
  {Amos}(1972)}]{Goddard1972}%
  \BibitemOpen
  \bibfield  {author} {\bibinfo {author} {\bibfnamefont {W.~A.}\ \bibnamefont
  {{Goddard III}}}\ and\ \bibinfo {author} {\bibfnamefont {A.}~\bibnamefont
  {Amos}},\ }\href@noop {} {\bibfield  {journal} {\bibinfo  {journal} {Chem.
  Phys. Lett.}\ }\textbf {\bibinfo {volume} {13}},\ \bibinfo {pages} {30}
  (\bibinfo {year} {1972})}\BibitemShut {NoStop}%
\bibitem [{\citenamefont {{Goddard III}}\ \emph {et~al.}(1973)\citenamefont
  {{Goddard III}}, \citenamefont {{Dunning Jr.}}, \citenamefont {Hunt},\ and\
  \citenamefont {Hay}}]{Goddard1973}%
  \BibitemOpen
  \bibfield  {author} {\bibinfo {author} {\bibfnamefont {W.~A.}\ \bibnamefont
  {{Goddard III}}}, \bibinfo {author} {\bibfnamefont {T.~H.}\ \bibnamefont
  {{Dunning Jr.}}}, \bibinfo {author} {\bibfnamefont {W.~J.}\ \bibnamefont
  {Hunt}}, \ and\ \bibinfo {author} {\bibfnamefont {P.~J.}\ \bibnamefont
  {Hay}},\ }\href@noop {} {\bibfield  {journal} {\bibinfo  {journal} {Acc.
  Chem. Res.}\ }\textbf {\bibinfo {volume} {6}},\ \bibinfo {pages} {368}
  (\bibinfo {year} {1973})}\BibitemShut {NoStop}%
\bibitem [{\citenamefont {Rassolov}(2002)}]{Rassolov-2002}%
  \BibitemOpen
  \bibfield  {author} {\bibinfo {author} {\bibfnamefont {V.~A.}\ \bibnamefont
  {Rassolov}},\ }\href@noop {} {\bibfield  {journal} {\bibinfo  {journal} {J.
  Chem. Phys.}\ }\textbf {\bibinfo {volume} {117}},\ \bibinfo {pages} {5978}
  (\bibinfo {year} {2002})}\BibitemShut {NoStop}%
\bibitem [{\citenamefont {Boguslawski}\ \emph {et~al.}(2014)\citenamefont
  {Boguslawski}, \citenamefont {Tecmer}, \citenamefont {Ayers}, \citenamefont
  {Bultinck}, \citenamefont {{De Baerdemacker}},\ and\ \citenamefont {{Van
  Neck}}}]{OO-AP1roG}%
  \BibitemOpen
  \bibfield  {author} {\bibinfo {author} {\bibfnamefont {K.}~\bibnamefont
  {Boguslawski}}, \bibinfo {author} {\bibfnamefont {P.}~\bibnamefont {Tecmer}},
  \bibinfo {author} {\bibfnamefont {P.~W.}\ \bibnamefont {Ayers}}, \bibinfo
  {author} {\bibfnamefont {P.}~\bibnamefont {Bultinck}}, \bibinfo {author}
  {\bibfnamefont {S.}~\bibnamefont {{De Baerdemacker}}}, \ and\ \bibinfo
  {author} {\bibfnamefont {D.}~\bibnamefont {{Van Neck}}},\ }\href@noop {}
  {\bibfield  {journal} {\bibinfo  {journal} {Phys. Rev. B}\ }\textbf {\bibinfo
  {volume} {submitted}} (\bibinfo {year} {2014})}\BibitemShut {NoStop}%
\bibitem [{\citenamefont {Helgaker}\ \emph {et~al.}(2000)\citenamefont
  {Helgaker}, \citenamefont {J{\o}rgensen},\ and\ \citenamefont
  {Olsen}}]{Helgaker_book}%
  \BibitemOpen
  \bibfield  {author} {\bibinfo {author} {\bibfnamefont {T.}~\bibnamefont
  {Helgaker}}, \bibinfo {author} {\bibfnamefont {P.}~\bibnamefont
  {J{\o}rgensen}}, \ and\ \bibinfo {author} {\bibfnamefont {J.}~\bibnamefont
  {Olsen}},\ }\href@noop {} {\emph {\bibinfo {title} {Molecular
  Electronic-Structure Theory}}}\ (\bibinfo  {publisher} {Wiley},\ \bibinfo
  {year} {2000})\BibitemShut {NoStop}%
\bibitem [{\citenamefont {Scuseria}\ and\ \citenamefont {{Schaefer
  III}}(1987)}]{Scuseria1987}%
  \BibitemOpen
  \bibfield  {author} {\bibinfo {author} {\bibfnamefont {G.~E.}\ \bibnamefont
  {Scuseria}}\ and\ \bibinfo {author} {\bibfnamefont {H.~F.}\ \bibnamefont
  {{Schaefer III}}},\ }\href@noop {} {\bibfield  {journal} {\bibinfo  {journal}
  {Chem. Phys. Lett.}\ }\textbf {\bibinfo {volume} {142}},\ \bibinfo {pages}
  {354} (\bibinfo {year} {1987})}\BibitemShut {NoStop}%
\bibitem [{\citenamefont {K\"{o}hn}\ and\ \citenamefont
  {Olsen}(2005)}]{Kohn2005}%
  \BibitemOpen
  \bibfield  {author} {\bibinfo {author} {\bibfnamefont {A.}~\bibnamefont
  {K\"{o}hn}}\ and\ \bibinfo {author} {\bibfnamefont {J.}~\bibnamefont
  {Olsen}},\ }\href@noop {} {\bibfield  {journal} {\bibinfo  {journal} {J.
  Chem. Phys.}\ }\textbf {\bibinfo {volume} {122}},\ \bibinfo {pages} {84116}
  (\bibinfo {year} {2005})}\BibitemShut {NoStop}%
\bibitem [{\citenamefont {Bozkaya}\ \emph {et~al.}(2011)\citenamefont
  {Bozkaya}, \citenamefont {Turney}, \citenamefont {Yamaguchi}, \citenamefont
  {Schaefer},\ and\ \citenamefont {Sherrill}}]{Ugur_2011}%
  \BibitemOpen
  \bibfield  {author} {\bibinfo {author} {\bibfnamefont {U.}~\bibnamefont
  {Bozkaya}}, \bibinfo {author} {\bibfnamefont {J.~M.}\ \bibnamefont {Turney}},
  \bibinfo {author} {\bibfnamefont {Y.}~\bibnamefont {Yamaguchi}}, \bibinfo
  {author} {\bibfnamefont {H.~F.}\ \bibnamefont {Schaefer}}, \ and\ \bibinfo
  {author} {\bibfnamefont {C.~D.}\ \bibnamefont {Sherrill}},\ }\href@noop {}
  {\bibfield  {journal} {\bibinfo  {journal} {J. Chem. Phys.}\ }\textbf
  {\bibinfo {volume} {135}},\ \bibinfo {pages} {104103} (\bibinfo {year}
  {2011})}\BibitemShut {NoStop}%
\bibitem [{\citenamefont {Limacher}\ \emph {et~al.}(2014)\citenamefont
  {Limacher}, \citenamefont {Kim}, \citenamefont {Ayers}, \citenamefont
  {Johnson}, \citenamefont {{De Baerdemacker}}, \citenamefont {Van~Neck},\ and\
  \citenamefont {Bultinck}}]{Piotrus_Mol-Phys}%
  \BibitemOpen
  \bibfield  {author} {\bibinfo {author} {\bibfnamefont {P.~A.}\ \bibnamefont
  {Limacher}}, \bibinfo {author} {\bibfnamefont {T.~D.}\ \bibnamefont {Kim}},
  \bibinfo {author} {\bibfnamefont {P.~W.}\ \bibnamefont {Ayers}}, \bibinfo
  {author} {\bibfnamefont {P.~A.}\ \bibnamefont {Johnson}}, \bibinfo {author}
  {\bibfnamefont {S.}~\bibnamefont {{De Baerdemacker}}}, \bibinfo {author}
  {\bibfnamefont {D.}~\bibnamefont {Van~Neck}}, \ and\ \bibinfo {author}
  {\bibfnamefont {P.}~\bibnamefont {Bultinck}},\ }\href@noop {} {\bibfield
  {journal} {\bibinfo  {journal} {Mol. Phys.}\ }\textbf {\bibinfo {volume}
  {112}},\ \bibinfo {pages} {853} (\bibinfo {year} {2014})}\BibitemShut
  {NoStop}%
\bibitem [{\citenamefont {Bytautas}\ \emph {et~al.}(2011)\citenamefont
  {Bytautas}, \citenamefont {Henderson}, \citenamefont {Jim\'{e}nez-Hoyos},
  \citenamefont {Ellis},\ and\ \citenamefont {Scuseria}}]{Bytautas2011}%
  \BibitemOpen
  \bibfield  {author} {\bibinfo {author} {\bibfnamefont {L.}~\bibnamefont
  {Bytautas}}, \bibinfo {author} {\bibfnamefont {T.~M.}\ \bibnamefont
  {Henderson}}, \bibinfo {author} {\bibfnamefont {C.~A.}\ \bibnamefont
  {Jim\'{e}nez-Hoyos}}, \bibinfo {author} {\bibfnamefont {J.~K.}\ \bibnamefont
  {Ellis}}, \ and\ \bibinfo {author} {\bibfnamefont {G.~E.}\ \bibnamefont
  {Scuseria}},\ }\href@noop {} {\bibfield  {journal} {\bibinfo  {journal} {J.
  Chem. Phys.}\ }\textbf {\bibinfo {volume} {135}},\ \bibinfo {pages} {044119}
  (\bibinfo {year} {2011})}\BibitemShut {NoStop}%
\bibitem [{\citenamefont {Alcoba}\ \emph {et~al.}(2013)\citenamefont {Alcoba},
  \citenamefont {Torre}, \citenamefont {Lain}, \citenamefont {Massaccesi},\
  and\ \citenamefont {Oña}}]{Alcoba2013}%
  \BibitemOpen
  \bibfield  {author} {\bibinfo {author} {\bibfnamefont {D.~R.}\ \bibnamefont
  {Alcoba}}, \bibinfo {author} {\bibfnamefont {A.}~\bibnamefont {Torre}},
  \bibinfo {author} {\bibfnamefont {L.}~\bibnamefont {Lain}}, \bibinfo {author}
  {\bibfnamefont {G.~E.}\ \bibnamefont {Massaccesi}}, \ and\ \bibinfo {author}
  {\bibfnamefont {O.~B.}\ \bibnamefont {Oña}},\ }\href@noop {} {\bibfield
  {journal} {\bibinfo  {journal} {J. Chem. Phys.}\ }\textbf {\bibinfo {volume}
  {139}},\ \bibinfo {pages} {084103} (\bibinfo {year} {2013})}\BibitemShut
  {NoStop}%
\bibitem [{\citenamefont {Giesbertz}(2014)}]{Giesbertz2014}%
  \BibitemOpen
  \bibfield  {author} {\bibinfo {author} {\bibfnamefont {K.}~\bibnamefont
  {Giesbertz}},\ }\href@noop {} {\bibfield  {journal} {\bibinfo  {journal}
  {Chem. Phys. Lett.}\ }\textbf {\bibinfo {volume} {591}},\ \bibinfo {pages}
  {220} (\bibinfo {year} {2014})}\BibitemShut {NoStop}%
\bibitem [{\citenamefont {Pian}\ and\ \citenamefont {Sharma}(1981)}]{Pian1981}%
  \BibitemOpen
  \bibfield  {author} {\bibinfo {author} {\bibfnamefont {J.}~\bibnamefont
  {Pian}}\ and\ \bibinfo {author} {\bibfnamefont {C.~S.}\ \bibnamefont
  {Sharma}},\ }\href@noop {} {\bibfield  {journal} {\bibinfo  {journal} {J.
  Phys. A: Math. Gen.}\ }\textbf {\bibinfo {volume} {14}},\ \bibinfo {pages}
  {1261} (\bibinfo {year} {1981})}\BibitemShut {NoStop}%
\bibitem [{\citenamefont {Crawford}\ and\ \citenamefont
  {Stanton}(2000)}]{Crawford2000}%
  \BibitemOpen
  \bibfield  {author} {\bibinfo {author} {\bibfnamefont {T.~D.}\ \bibnamefont
  {Crawford}}\ and\ \bibinfo {author} {\bibfnamefont {J.~F.}\ \bibnamefont
  {Stanton}},\ }\href@noop {} {\bibfield  {journal} {\bibinfo  {journal} {J.
  Chem. Phys.}\ }\textbf {\bibinfo {volume} {112}},\ \bibinfo {pages} {7873}
  (\bibinfo {year} {2000})}\BibitemShut {NoStop}%
\bibitem [{\citenamefont {Wouters}\ \emph {et~al.}(2014)\citenamefont
  {Wouters}, \citenamefont {Poelmans}, \citenamefont {Ayers},\ and\
  \citenamefont {{Van Neck}}}]{CheMPS2}%
  \BibitemOpen
  \bibfield  {author} {\bibinfo {author} {\bibfnamefont {S.}~\bibnamefont
  {Wouters}}, \bibinfo {author} {\bibfnamefont {W.}~\bibnamefont {Poelmans}},
  \bibinfo {author} {\bibfnamefont {P.~W.}\ \bibnamefont {Ayers}}, \ and\
  \bibinfo {author} {\bibfnamefont {D.}~\bibnamefont {{Van Neck}}},\
  }\href@noop {} {\bibfield  {journal} {\bibinfo  {journal} {Comput. Phys.
  Comm.}\ }\textbf {\bibinfo {volume} {XX}},\ \bibinfo {pages}
  {DOI:10.1016/j.cpc.2014.01.019} (\bibinfo {year} {2014})}\BibitemShut
  {NoStop}%
\bibitem [{\citenamefont {Evangelista}(2011)}]{Evangelista_2011}%
  \BibitemOpen
  \bibfield  {author} {\bibinfo {author} {\bibfnamefont {F.~A.}\ \bibnamefont
  {Evangelista}},\ }\href@noop {} {\bibfield  {journal} {\bibinfo  {journal}
  {J. Chem. Phys.}\ }\textbf {\bibinfo {volume} {134}},\ \bibinfo {pages}
  {224102} (\bibinfo {year} {2011})}\BibitemShut {NoStop}%
\bibitem [{\citenamefont {Peterson}(1995)}]{Peterson1995}%
  \BibitemOpen
  \bibfield  {author} {\bibinfo {author} {\bibfnamefont {K.~A.}\ \bibnamefont
  {Peterson}},\ }\href@noop {} {\bibfield  {journal} {\bibinfo  {journal} {J.
  Chem. Phys.}\ }\textbf {\bibinfo {volume} {102}},\ \bibinfo {pages} {262}
  (\bibinfo {year} {1995})}\BibitemShut {NoStop}%
\bibitem [{\citenamefont {Legeza}\ \emph {et~al.}(2009)\citenamefont {Legeza},
  \citenamefont {R\"{o}der},\ and\ \citenamefont {Hess}}]{LegezaLiF}%
  \BibitemOpen
  \bibfield  {author} {\bibinfo {author} {\bibfnamefont {O.}~\bibnamefont
  {Legeza}}, \bibinfo {author} {\bibfnamefont {J.}~\bibnamefont {R\"{o}der}}, \
  and\ \bibinfo {author} {\bibfnamefont {B.~A.}\ \bibnamefont {Hess}},\
  }\href@noop {} {\bibfield  {journal} {\bibinfo  {journal} {Mol. Phys.}\
  }\textbf {\bibinfo {volume} {101}},\ \bibinfo {pages} {37} (\bibinfo {year}
  {2009})}\BibitemShut {NoStop}%
\bibitem [{\citenamefont {Varandas}(2009)}]{Varandas2009}%
  \BibitemOpen
  \bibfield  {author} {\bibinfo {author} {\bibfnamefont {A.~J.~C.}\
  \bibnamefont {Varandas}},\ }\href@noop {} {\bibfield  {journal} {\bibinfo
  {journal} {J. Chem. Phys.}\ }\textbf {\bibinfo {volume} {131}},\ \bibinfo
  {pages} {124128} (\bibinfo {year} {2009})}\BibitemShut {NoStop}%
\bibitem [{Hor()}]{Horton13}%
  \BibitemOpen
  \href@noop {} {}\bibinfo {note} {Horton 1.2.0 2013, written by T.
  Verstraelen, S. Vandenbrande, M. Chan, F. H. Zadeh, C. Gonzalez, K.
  Boguslawski, P. Tecmer, P. A. Limacher, A. Malek (see {\tt
  http://theochem.github.com/horton/})}\BibitemShut {NoStop}%
\bibitem [{\citenamefont {Coxon}(1992)}]{Coxon_1992}%
  \BibitemOpen
  \bibfield  {author} {\bibinfo {author} {\bibfnamefont {J.~A.}\ \bibnamefont
  {Coxon}},\ }\href@noop {} {\bibfield  {journal} {\bibinfo  {journal} {J. Mol.
  Spectrosc.}\ }\textbf {\bibinfo {volume} {282}},\ \bibinfo {pages} {274}
  (\bibinfo {year} {1992})}\BibitemShut {NoStop}%
\bibitem [{\citenamefont {Abramowitz}\ and\ \citenamefont
  {Stegun}(1970)}]{Abramowitz}%
  \BibitemOpen
  \bibfield  {author} {\bibinfo {author} {\bibfnamefont {M.}~\bibnamefont
  {Abramowitz}}\ and\ \bibinfo {author} {\bibfnamefont {I.~A.}\ \bibnamefont
  {Stegun}},\ }\href@noop {} {\emph {\bibinfo {title} {Handbook of Mathematical
  Functions with Formulas, Graphs, and Mathematical Tables}}}\ (\bibinfo
  {publisher} {Dover},\ \bibinfo {year} {1970})\BibitemShut {NoStop}%
\bibitem [{\citenamefont {Dunning}(1989)}]{dunning_b}%
  \BibitemOpen
  \bibfield  {author} {\bibinfo {author} {\bibfnamefont {T.~H.}\ \bibnamefont
  {Dunning}},\ }\href@noop {} {\bibfield  {journal} {\bibinfo  {journal}
  {J.~Chem.~Phys.}\ }\textbf {\bibinfo {volume} {90}},\ \bibinfo {pages} {1007}
  (\bibinfo {year} {1989})}\BibitemShut {NoStop}%
\bibitem [{\citenamefont {Feller}(1996)}]{emsl-basis-1}%
  \BibitemOpen
  \bibfield  {author} {\bibinfo {author} {\bibfnamefont {D.}~\bibnamefont
  {Feller}},\ }\href@noop {} {\bibfield  {journal} {\bibinfo  {journal} {J.
  Comp. Chem.}\ }\textbf {\bibinfo {volume} {17}},\ \bibinfo {pages} {1571}
  (\bibinfo {year} {1996})}\BibitemShut {NoStop}%
\bibitem [{\citenamefont {Didier}\ \emph {et~al.}(2007)\citenamefont {Didier},
  \citenamefont {Elsethagen}, \citenamefont {T.~Sun}, \citenamefont {Chase},
  \citenamefont {Li},\ and\ \citenamefont {Windus}}]{emsl-basis-2}%
  \BibitemOpen
  \bibfield  {author} {\bibinfo {author} {\bibfnamefont {K.}~\bibnamefont
  {Didier}}, \bibinfo {author} {\bibfnamefont {B.}~\bibnamefont {Elsethagen}},
  \bibinfo {author} {\bibfnamefont {L.~G.}\ \bibnamefont {T.~Sun}}, \bibinfo
  {author} {\bibfnamefont {V.}~\bibnamefont {Chase}}, \bibinfo {author}
  {\bibfnamefont {J.}~\bibnamefont {Li}}, \ and\ \bibinfo {author}
  {\bibfnamefont {T.~L.}\ \bibnamefont {Windus}},\ }\href@noop {} {\bibfield
  {journal} {\bibinfo  {journal} {J. Chem. Inf. Model.}\ }\textbf {\bibinfo
  {volume} {47}},\ \bibinfo {pages} {1045} (\bibinfo {year}
  {2007})}\BibitemShut {NoStop}%
\bibitem [{\citenamefont {Woon}\ and\ \citenamefont
  {Dunning}(1995)}]{cc-pCVDZ}%
  \BibitemOpen
  \bibfield  {author} {\bibinfo {author} {\bibfnamefont {D.~E.}\ \bibnamefont
  {Woon}}\ and\ \bibinfo {author} {\bibfnamefont {T.~H.}\ \bibnamefont
  {Dunning}},\ }\href@noop {} {\bibfield  {journal} {\bibinfo  {journal} {J.
  Chem. Phys.}\ }\textbf {\bibinfo {volume} {103}},\ \bibinfo {pages} {4572}
  (\bibinfo {year} {1995})}\BibitemShut {NoStop}%
\bibitem [{\citenamefont {Hehre}\ \emph {et~al.}(1969)\citenamefont {Hehre},
  \citenamefont {Stewart},\ and\ \citenamefont {Pople}}]{STO-6G-H-Ne}%
  \BibitemOpen
  \bibfield  {author} {\bibinfo {author} {\bibfnamefont {W.~J.}\ \bibnamefont
  {Hehre}}, \bibinfo {author} {\bibfnamefont {R.~F.}\ \bibnamefont {Stewart}},
  \ and\ \bibinfo {author} {\bibfnamefont {J.~A.}\ \bibnamefont {Pople}},\
  }\href@noop {} {\bibfield  {journal} {\bibinfo  {journal} {J. Chem. Phys.}\
  }\textbf {\bibinfo {volume} {51}},\ \bibinfo {pages} {2657} (\bibinfo {year}
  {1969})}\BibitemShut {NoStop}%
\bibitem [{\citenamefont {Hachmann}\ \emph {et~al.}(2006)\citenamefont
  {Hachmann}, \citenamefont {Cardoen},\ and\ \citenamefont
  {Chan}}]{Hachmann_H50}%
  \BibitemOpen
  \bibfield  {author} {\bibinfo {author} {\bibfnamefont {J.}~\bibnamefont
  {Hachmann}}, \bibinfo {author} {\bibfnamefont {W.}~\bibnamefont {Cardoen}}, \
  and\ \bibinfo {author} {\bibfnamefont {G.~K.-L.}\ \bibnamefont {Chan}},\
  }\href@noop {} {\bibfield  {journal} {\bibinfo  {journal} {J. Chem. Phys.}\
  }\textbf {\bibinfo {volume} {125}},\ \bibinfo {pages} {144101} (\bibinfo
  {year} {2006})}\BibitemShut {NoStop}%
\bibitem [{Note1()}]{Note1}%
  \BibitemOpen
  \bibinfo {note} {We should note that all calculations are performed in C$_1$
  symmetry (see computational details). Therefore, molecular orbitals cannot be
  labeled according to an irreducible representation of the molecular point
  group. The term 'symmetry-adapted' is used to emphasize that the optimized
  molecular orbitals are not symmetry-broken (\protect \emph {e.g.}, localized,
  hybrid, etc.); this does not imply that the molecular orbitals transform as
  an irreducible representation of the molecular point group.}\BibitemShut
  {Stop}%
\bibitem [{\citenamefont {McLean}\ \emph {et~al.}(1985)\citenamefont {McLean},
  \citenamefont {{Lengsfield III}}, \citenamefont {Pacansky},\ and\
  \citenamefont {Ellinger}}]{McLean1985}%
  \BibitemOpen
  \bibfield  {author} {\bibinfo {author} {\bibfnamefont {A.~D.}\ \bibnamefont
  {McLean}}, \bibinfo {author} {\bibfnamefont {B.~H.}\ \bibnamefont
  {{Lengsfield III}}}, \bibinfo {author} {\bibfnamefont {J.}~\bibnamefont
  {Pacansky}}, \ and\ \bibinfo {author} {\bibfnamefont {Y.}~\bibnamefont
  {Ellinger}},\ }\href@noop {} {\bibfield  {journal} {\bibinfo  {journal} {J.
  Chem. Phys.}\ }\textbf {\bibinfo {volume} {83}},\ \bibinfo {pages} {3567}
  (\bibinfo {year} {1985})}\BibitemShut {NoStop}%
\bibitem [{Jmo()}]{Jmol}%
  \BibitemOpen
  \href@noop {} {}\bibinfo {note} {Jmol: An Open-Source Java Viewer for
  Chemical Structures in 3D. {\tt http://www.jmol.org/}}\BibitemShut {NoStop}%
\bibitem [{\citenamefont {Tsuchimochi}\ and\ \citenamefont
  {Scuseria}(2009)}]{Gustavo_H50}%
  \BibitemOpen
  \bibfield  {author} {\bibinfo {author} {\bibfnamefont {T.}~\bibnamefont
  {Tsuchimochi}}\ and\ \bibinfo {author} {\bibfnamefont {G.~E.}\ \bibnamefont
  {Scuseria}},\ }\href@noop {} {\bibfield  {journal} {\bibinfo  {journal} {J.
  Chem. Phys.}\ }\textbf {\bibinfo {volume} {131}},\ \bibinfo {pages} {121102}
  (\bibinfo {year} {2009})}\BibitemShut {NoStop}%
\bibitem [{\citenamefont {Stella}\ \emph {et~al.}(2011)\citenamefont {Stella},
  \citenamefont {Attaccalite}, \citenamefont {Sorella},\ and\ \citenamefont
  {Rubio}}]{Stella_H50}%
  \BibitemOpen
  \bibfield  {author} {\bibinfo {author} {\bibfnamefont {L.}~\bibnamefont
  {Stella}}, \bibinfo {author} {\bibfnamefont {C.}~\bibnamefont {Attaccalite}},
  \bibinfo {author} {\bibfnamefont {S.}~\bibnamefont {Sorella}}, \ and\
  \bibinfo {author} {\bibfnamefont {A.}~\bibnamefont {Rubio}},\ }\href@noop {}
  {\bibfield  {journal} {\bibinfo  {journal} {Phys. Rev. B}\ }\textbf {\bibinfo
  {volume} {84}},\ \bibinfo {pages} {245117} (\bibinfo {year}
  {2011})}\BibitemShut {NoStop}%
\bibitem [{\citenamefont {Lin}\ \emph {et~al.}(2011)\citenamefont {Lin},
  \citenamefont {Marianetti}, \citenamefont {Millis},\ and\ \citenamefont
  {Reichman}}]{DMFT_H50}%
  \BibitemOpen
  \bibfield  {author} {\bibinfo {author} {\bibfnamefont {N.}~\bibnamefont
  {Lin}}, \bibinfo {author} {\bibfnamefont {C.~A.}\ \bibnamefont {Marianetti}},
  \bibinfo {author} {\bibfnamefont {A.~J.}\ \bibnamefont {Millis}}, \ and\
  \bibinfo {author} {\bibfnamefont {D.~R.}\ \bibnamefont {Reichman}},\
  }\href@noop {} {\bibfield  {journal} {\bibinfo  {journal} {Phys. Rev. Lett.}\
  }\textbf {\bibinfo {volume} {106}},\ \bibinfo {pages} {096402} (\bibinfo
  {year} {2011})}\BibitemShut {NoStop}%
\end{thebibliography}%
\end{document}